\documentclass[conference]{IEEEtran}
\usepackage{pifont}
\usepackage{graphics}
\usepackage{amsmath}
\usepackage{amssymb}
\usepackage{url}
\usepackage{multirow}
\usepackage[table,xcdraw]{xcolor}
\usepackage{colortbl}

\usepackage{fancyhdr}
\usepackage{kantlipsum}

\usepackage{eso-pic}

\usepackage[linesnumbered,ruled,vlined]{algorithm2e}

\SetCommentSty{mycommfont}
\usepackage{amsmath}

\DeclareMathOperator*{\argmax}{\arg\!\max}
\newcommand{\INV}{\mathbb{INV}}
\newcommand{\SIM}{\mathbb{SIM}}
\newcommand{\ENDISCO}{{\tt EnDisCo}}
\newcommand{\MEDOC}{{\tt MeDOC}}
\newcommand{\LOU}{{\tt Louvain}}
\newcommand{\CNM}{{\tt CNM}}
\newcommand{\INF}{{\tt InfoMap}}
\newcommand{\WT}{{\tt WalkTrap}}
\newcommand{\LP}{{\tt LabelPr}}
\newcommand{\CC}{{\tt ConsCl}}
\newcommand{\FG}{{\tt FstGrdy}}
\newcommand{\OSLOM}{{\tt OSLOM}}
\newcommand{\EAGLE}{{\tt EAGLE}}
\newcommand{\COPRA}{{\tt COPRA}} 
\newcommand{\SLPA}{{\tt SLPA}}
\newcommand{\MOSES}{{\tt MOSES}}
\newcommand{\BIGCLAM}{{\tt  BIGCLAM}}
\newcommand{\nop}[1]{{}}

\ifCLASSINFOpdf
\else
\fi

\begin{document}
\AddToShipoutPictureBG*{%
  \AtPageUpperLeft{%
    \setlength\unitlength{1in}%
    \hspace*{\dimexpr0.5\paperwidth\relax}
    \makebox(0,-0.75)[c]{\textbf{2016 IEEE/ACM International Conference on Advances in Social Networks Analysis
and Mining (ASONAM)}}%
}}

\title{Ensemble-Based Algorithms to Detect Disjoint and Overlapping Communities in Networks}

\author{\IEEEauthorblockN{Tanmoy Chakraborty}
\IEEEauthorblockA{Dept. of Computer Science\\
University of Maryland\\
College Park, MD 20742 \\
Email: tanchak@umiacs.umd.edu}
\and
\IEEEauthorblockN{Noseong Park}
\IEEEauthorblockA{Dept. of Computer Science\\
University of Maryland\\
College Park, MD 20742 \\
Email: npark@cs.umd.edu}
\and
\IEEEauthorblockN{V.S. Subrahmanian }
\IEEEauthorblockA{Dept. of Computer Science\\
University of Maryland\\
College Park, MD 20742 \\
Email: vs@umiacs.umd.edu}}

\IEEEoverridecommandlockouts
\IEEEpubid{\makebox[\columnwidth]{IEEE/ACM ASONAM 2016, August 18-21, 2016, San Francisco, CA, USA
978-1-5090-2846-7/16/\$31.00~
\copyright2016
IEEE \hfill}
}

\maketitle

\begin{abstract}
Given a set ${\cal AL}$ of community detection algorithms and a graph $G$ as inputs, we propose two ensemble methods 
\ENDISCO~ and \MEDOC~that (respectively) identify disjoint and overlapping communities in $G$. 
 \ENDISCO~transforms a  graph into a latent feature space by leveraging multiple base solutions and discovers  disjoint community structure. \MEDOC~groups similar base communities into a \emph{meta-community} and detects both disjoint and overlapping community structures. Experiments are conducted at different scales on both synthetically generated networks as well as on several real-world networks for which the underlying ground-truth community structure is available. Our extensive experiments show that both algorithms outperform state-of-the-art non-ensemble algorithms by  a significant margin. Moreover, we compare \ENDISCO~and \MEDOC~with a recent ensemble method for disjoint community detection and show that our approaches achieve superior performance. To the best of our knowledge, \MEDOC~is the first {\em ensemble approach} for overlapping community detection.
\end{abstract}

\IEEEpeerreviewmaketitle

\section{Introduction}
\begin{figure}[!h]
 \centering
 \scalebox{0.17}{
 \includegraphics{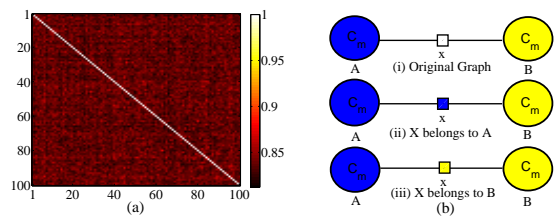}}
 \caption{(a) Similarity between pair-wise community structures (based on NMI \cite{danon2005ccs}) after running Infomap algorithm on $100$ different vertex orderings of the Football network. (b) A schematic network consisting of two cliques $A$ and $B$ of size $m$ (representing two communities) connected by a bridging vertex $x$. Assigning vertex $x$ to either $A$ or $B$ yields the same value of the optimization metrics (such as modularity \cite{Newman:2006:}).}\label{example}
 \vspace{-3mm}
\end{figure}

Community detection (CD) has found applications in social, biological, business, and other kinds of networks. 
However, CD algorithms suffer from various flaws -- (i) Most existing CD algorithms are heavily dependent on {\em vertex ordering} \cite{good2010}, yielding completely different community structures when the same network is processed in a different order. For example, Figure \ref{example}(a) shows  dissimilar community structures  after running \INF~ \cite{Rosvall29012008} on $100$ different vertex orderings of the Football network \cite{Clauset2004}. (ii) Most optimization algorithms may produce multiple community structures with the same ``optimal'' value of the objective function. For instance, in  Figure \ref{example}(b), assigning vertex $x$ to either $A$ or $B$ results in the same modularity score \cite{Newman:2006:}. (iii)  Different CD algorithms detect communities in different ways, e.g., as dense groups internally \cite{newman03fast}, or groups with sparse connections externally \cite{Fortunato201075}. 
It is therefore natural to think of an ensemble approach in which the strengths of different CD algorithms may help overcome the weaknesses of any specific CD algorithm. Some preliminary attempts have been made by \cite{DahlinS13,lanc12consensus}\footnote{Note that
ensemble approaches have proved successful in clustering and classification \cite{Xu:2005}.}.

\noindent{\bf Contributions.} In this paper, we design two ensemble CD algorithms.
  The \ENDISCO~algorithm runs multiple ``base'' CD algorithms using a variety of vertex orderings to derive a first set of communities. We then consider the memberships of vertices obtained from base CD algorithms as features and derive a latent network using pair-wise similarity of vertices. The final disjoint community structure is obtained by running any CD algorithm again on the latent network. The \MEDOC~ algorithm leverages the fact that many communities returned by base algorithms are redundant and can therefore be grouped into 
``meta-communities'' to avoid unnecessary computation. We use meta-communities to build an association matrix, where each entry indicates the probability of a vertex belonging to a meta-community. Finally, we obtain  both disjoint and overlapping community structures via post-processing on the association matrix. 
\emph{To the best of our knowledge, we are the first to propose (i) an ensemble framework for overlapping community detection, and (ii) an overlapping CD algorithm that leverages disjoint community information.} We run experiments to identify the best parameter settings for \ENDISCO~and \MEDOC.
 Experiments on both synthetic and real-world networks show that our algorithms outperform both the state-of-the-art non-ensemble based methods \cite{blondel2008,Rosvall29012008,Leskovec} and a recently proposed ensemble approach \cite{lanc12consensus} by a significant margin\footnote{We report $p$-values for all our experiments to show statistical significance.}. We also show that our ensemble approaches  reduce the effect of vertex ordering. 

\IEEEpubidadjcol

\noindent\textbf{Note:} We use the term ``community structure'' to indicate the result (set of ``communities'') returned by an algorithm. Each community is a set of vertices.


\section{Related Work}
There  has been a great deal of work on clustering data using ensemble approaches (see \cite{Xu:2005} for a review). 
However, when it comes to clustering vertices in \emph{networks}, ensemble approaches have been relatively scarce\footnote{See the survey \cite{Fortunato201075} for various community detection algorithms.}. Dahlin and Svenson \cite{DahlinS13} were the first to propose an instance-based ensemble CD algorithm for networks which fuses different community structures into a final representation.  A few methods addressed the utility of merging several community structures \cite{raghavan-2007}. A new ensemble scheme called CGGC was proposed to maximize modularity  \cite{OvelgonneG12}. Kanawati proposed YASCA, an ensemble approach to different network partitions derived from ego-centered communities computed for each selected seed \cite{Kanawati2014}. He further emphasized the quality and diversity of outputs obtained from the base algorithms for ensemble selection \cite{Kanawati2015}.

A   ``consensus clustering'' \cite{lanc12consensus} approach was recently proposed  which leverages a {\em consensus matrix} to produce a disjoint community structure which outperformed previous approaches.  Our work differs from this approach in at least three significant ways: (i) they measure the number of times two vertices are assigned to the same community, thus ignoring the global similarity of vertices; whereas we capture the global similarity  by representing the network within a feature space and grouping redundant base communities into meta communities; (ii) they either take multiple algorithms or run a particular algorithm multiple times for generating an ensemble matrix, whereas we consider both options; (iii) we are the first to show how aggregating multiple disjoint base communities can lead to discover both disjoint and overlapping community structures simultaneously. We show experimentally that \ENDISCO~beats consensus clustering.


\begin{algorithm}\label{insimul}\small
\caption{\ENDISCO: {\bf En}semble-based {\bf Dis}joint {\bf Co}mmunity Detection }
\KwData{Graph $G(V,E)$; \\Base algorithms ${\cal AL}=\{Al_m\}_{m=1}^M$; \\$K$: Number of iterations; \\$\INV(.,.)$: Involvement function;\\ $\SIM(.,.)$: Similarity function between two vectors; \\$RAlgo$: Algorithm for re-clustering}
\KwResult{Disjoint community structure $\mathbb{DC}$}

$\Gamma=\phi$   \hfill \tcp{Set of all base community structures} 

\tcp{{\color{blue} Generating base partitions}}
\For{each algorithm $Al_m\in {\cal AL}$}{
Run $Al_m$ on $G$ for $K$ different vertex orderings and obtain $K$ community structures, denoted by the set $\Gamma_m$; each community structure $\mathbb{C}_m^k \in \Gamma_m$ is of different size and indicated by $\mathbb{C}_m^k=\{C_m^{1k},...,C_m^{ak}\}$;\label{algo1:ensemble}\\
$\Gamma=\Gamma \cup \Gamma_m$;
}

\For{each $v$ in $V$}{\label{algo1:v}
 $F(v)=\phi$; \hfill \tcp{Feature vector of $v$}
 $D_v=0$; \hfill \tcp{Max distance of $v$ to any community}
 $Clu=0$; \hfill \tcp{Total no of communities}
 \tcp{{\color{blue} Constructing ensemble matrix}}
 \For{each $\Gamma_m \in \Gamma$}{\label{gammas} 
     \For{each $\mathbb{C}_m^k\in \Gamma_m$}{
      \For{each $C\in \mathbb{C}_m^k$}{\label{algo1:c}
       Compute $d_v^C=1-\INV(v,C)$;\label{algo1:dist} \\
       $F(v)=F(v) \cup d_v^C$;\label{algo1:fea}\\
       \If{$ d_v^C \geq D_v $}{
	    $D_v=d_v^C$;\label{algo1:D}
        }
        $Clu=Clu+1$;\label{algo1:cl}
       }      
   }
 }\label{gammae} 
 $P(v)=\phi$;\\
 \For{each $F_i(v) \in F(v)$}{\label{algo1:pre}
     \tcp{Posterior probability of $v$ in $C_i^k$}
     Compute $P(C_i|v) =\frac{D_v-F_i(v) +1}{Clu.D_v + Clu -\sum_{k=1}^{Clu} F_k(v)}$;\label{algo1:prob}\\
     $P(v)=P(v)\cup P(C_i|v)$;\label{algo1:post}
   } 
 }
Build an ensemble matrix $\mathbb{M}_{|V|\times|V|}$, where $\forall u,v\in V;\ \mathbb{M}(u,v)$=$\SIM(P(u),P(v))$;\label{algo1:sim}\\
\tcp{{\color{blue} Re-clustering the vertices from $M$}}
Run $RAlgo$ for {\em re-clustering vertices} from $M$ and discover a disjoint community structure $\mathbb{DC}$;\label{algo1:assign} \\
\Return $\mathbb{DC}$
\end{algorithm}

\section{\ENDISCO: Ensemble-based Disjoint Community Detection}
\ENDISCO~({\bf En}semble-based {\bf Dis}joint {\bf Co}mmunity Detection) starts by first using different CD algorithms to identify different community structures. Second, an ``involvement'' function is  
used to measure the extent to which a vertex is involved with  a given community, which in turn sets the  posterior probabilities of each vertex belonging to different communities. Third,
\ENDISCO~transforms the network into a feature space. Fourth, an ensemble matrix is constructed by measuring the pair-wise similarity of vertices. Finally, we apply any standard CD algorithm on the ensemble matrix and discover the final disjoint community structure.

\subsection{Algorithmic Description}
 \ENDISCO~ follows three fundamental steps (a pseudo-code is shown in Algorithm \ref{insimul}):
 
\noindent{\bf (i) Generating base partitions.}  Given a network $G=(V,E)$ and a set ${\cal AL}=\{Al_m\}_{m=1}^M$ of $M$ different base CD algorithms, \ENDISCO~runs each algorithm $Al_m$ on $K$ different vertex orderings (randomly selected) of $G$. This generates a set of $K$ different community structures denoted $\Gamma_m=\{\mathbb{C}_m^k\}_{k=1}^K$, where each community structure $\mathbb{C}_m^k=\{C_m^{1k},\cdots,C_m^{ak}\}$ constitutes a specific partitioning of vertices in $G$, and each $\mathbb{C}_m^k$ might be of different size (Step \ref{algo1:ensemble}). 

\noindent{\bf (ii) Constructing ensemble matrix.} Given a $\Gamma_m$, we then compute the extent of $v$'s involvement in each community $C$ in $\mathbb{C}_m^k$ via an ``involvement'' function $\INV(v,C)$ (Step \ref{algo1:dist}). Possible definitions of $\INV$ are given in Section \ref{algo1:Paramater}. For  each vertex $v$, we construct a feature vector $F(v)$ whose elements indicate the distance  of $v$ (measured by $1-\INV$) from each community obtained from different runs of the base algorithms (Step \ref{algo1:fea}). The size of $F(v)$ is same as the number of communities $Clu$ in $\Gamma$ (approx. $\bar{a}MK$, where $\bar{a}$ is the average size of a base community structure). Let $D_v$ be the largest distance of $v$ from any community in the sets in $\Gamma$ (i.e., $D_v=\max_i F_i(v)$ in Step \ref{algo1:D}). We define the conditional probability of $v$ belonging to community $C_i$  (Step \ref{algo1:prob}) as:
\begin{equation}\small
 P(C_i|v) =\frac{D_v-F_i(v) +1}{Clu.D_v + Clu -\sum_{k=1}^{Clu} F_k(v)}
\end{equation}
The numerator ensures that the greater the distance $F_i(v)$ of $v$ from community $C_i$, the less likely $v$ is to be in community $C_i$. 
The normalization factor in the denominator ensures that $\sum_{k=1}^{Clu} P(C_i|v)=1$. Add-one smoothing in the numerator allows a non-zero probability to be assigned to all $C_i$s, especially for $C_{\hat{k}}$ such that $\hat{k}=\argmax\limits_{k} F_k(v)$. 

The set of posterior probabilities of $v$ is: $P(v)=\{ P(C_k|v) \}_{k=1}^ {Clu}$ (Step \ref{algo1:post}), which in turn transforms a vertex into a point in a multi-dimensional feature space. Finally we construct an ensemble matrix $M$ whose entry $M(u,v)$ is the similarity (obtained from a function $\SIM$ whose possible definitions are given in Section \ref{algo1:Paramater}) between the feature vectors of $u$ and $v$ (Step \ref{algo1:sim}). The ensemble matrix ensures that the more communities a pair of vertices share the more likely they are connected in the network \cite{Leskovec}. 

\noindent{\bf (iii) Discovering final community structure.}  In Step \ref{algo1:assign} we use a CD algorithm $RAlgo$ to re-cluster the vertices from $M$ and discover the final disjoint community structure (Step \ref{algo1:assign}).

\subsection{Parameter Selection}\label{algo1:Paramater}
\noindent $\bullet$ {\bf Involvement Function ($\INV$):} We use two functions to measure the involvement of a vertex $v$ in a community $C$: (i) {\em Restricted Closeness Centrality} ($RCC$): This is the inverse of the average shortest-path distance from the vertex $v$ to the vertices in community $C$, i.e., $RCC(v,C)=\frac{|C|}{\sum_{u\in C} dist(v,u)}$; (ii) {\em Inverse Distance from Centroid} ({\em IDC}): we first identify the vertex with highest closeness centrality (w.r.t. the induced subgraph of $C$) in community $C$, mark it as the centroid of $C$ (denoted by $u_c$), and then measure  the involvement of $v$ as the inverse of the shortest-path distance between $v$ and $u_c$, i.e.,  $IDC(v,C)=\frac{1}{dist(v,u_c)}$.

\noindent $\bullet$ {\bf Similarity Function ($\SIM$):} We consider cosine similarity ($COS$) and Chebyshev distance ($CHE$) (essentially, $1-CHE$) to measure the similarity between two vectors.

\noindent $\bullet$ {\bf Algorithm for Re-clustering ($RAlgo$):} we consider each base CD algorithm as the one to re-cluster the vertices from the ensemble matrix. The idea is to show that a non-ensemble CD algorithm can perform even better when considering the ensemble matrix of network $G$ than the adjacency matrix of $G$.  However, one can use any CD algorithm in this step to detect the community structure. We will show the effect of different algorithms used in this step in Section \ref{sec:impact}.

\noindent $\bullet$ {\bf Number of Iterations ($K$):} Instead of fixing a hard value, we set $K$ to be dependent on the number of vertices $|V|$ in the network. We vary $K$ from $0.01$ to $0.50$ (with step $0.05$) of $|V|$ and confirm that for most of the networks, the accuracy of the algorithm converges at $K=0.2|V|$ (Figures \ref{parameter_dis}(c) and \ref{parameter_dis}(f)), and therefore we set $K=0.2|V|$
in our experiments. 

\subsection{Complexity Analysis}\label{algo1:complexity}

Suppose $N=|V|$ is the number of vertices in the network, $M$ is the  number of base algorithms and $K$ is the number of vertex orderings. Further suppose $\bar{a}$ is the average size of the community structure. Then the loop in Step \ref{algo1:v} of Algorithm 1  would iterate $\bar{a}NMK$ times (where $M,K\ll N$). The construction of the ensemble matrix in Step \ref{algo1:sim} would take $\mathcal{O}(N^2)$. 
Graph partitioning is NP-hard even to find a solution with guaranteed approximation bounds --- however, heuristics such as the famous Kernighan-Lin algorithm take $O(N^2\cdot \mbox{log}(N))$ time.

\vspace{-3mm}

\begin{algorithm}\label{meclud}\small
\caption{{\MEDOC}: A {\bf Me}ta Clustering based {\bf D}isjoint and {\bf O}verlapping {\bf C}ommunity Detection}
\KwData{Graph $G(V,E)$;\\ Base algorithms ${\cal AL}=\{Al_m\}_{m=1}^ M$;\\ $K$: Number of iterations; \\$W(.,.)$: Matching between pair-wise communities;\\ $RAlgo$: Algorithm for re-clustering; \\${\cal F}(.,.)$: vertex-to-community association; \\ $\tau$: threshold for overlapping community detection}
\KwResult{Disjoint ($\mathbb{DC}$) and overlapping ($\mathbb{OC}$) community structures}
\tcp{{\color{blue} Constructing multipartite network}}
\For{$Al_m$ in ${\cal AL}$}{
Run $Al_m$ on $G$ for $K$ different vertex orderings and obtain $K$  community structures, denoted by the set $\Gamma_m=\{\mathbb{C}_m^k\}_{k=1}^K$; each community structure $\mathbb{C}_m^k \in \Gamma_m$ may be of different size and is denoted by $\mathbb{C}_m^k=\{C_m^{1k},...,C_i^{ak}\}$; \label{algo2:perm}}
Construct a $P$-partite graph $GP$ (where $P=M.K$)  consisting of $M.K$ partitions, each corresponding to each community structure obtained in Step 2: vertices in partition $m^k$ are communities in $\mathbb{C}_m^k$ and edges are drawn between two pair-wise vertices (communities) $C_m^{ik}$ and $C_n^{jk}$ with the edge weight $W(C_m^{ik},C_n^{jk'})$;\label{algo2:cons}\\
\tcp{{\color{blue} Re-clustering the multipartite network}}
Run $RAlgo$ to re-cluster vertices in $GP$ and discover a meta-community structure, $\mathbb{C}_{GP}=\{C_{GP}^{l}\}_{l=1}^L$;\label{algo2:run}\\
\tcp{{\color{blue} Constructing an association matrix}}
Construct an association matrix $\mathbb{A}_{|V|\times L}$, where ${\mathbb A}(v,l)={\cal F}(v,C_{GP}^l)$, indicating the association of vertex $v$ to a meta-community $C_{GP}^l$;\label{algo2:asso}\\
\tcp{{\color{blue} Discovering final community structure}}
Each row in $\mathbb{A}$ indicates the membership probabilities of the corresponding vertex in $L$ meta-communities;\\
To get $\mathbb{DC}$, we assign a vertex $v$ to community $C^* = \argmax\limits_{C} \ \mathbb{A} (v,C) $;\label{algo2:dc}\\
To get $\mathbb{OC}$, we assign a vertex $v$ to a set of communities $C_v^*$ so that $\forall C\in C_v^*: \mathbb{A} (v,C) \geq \tau$;\label{algo2:oc}\\
\Return $\mathbb{DC}$, $\mathbb{OC}$
\end{algorithm}

\vspace{-5mm}

\section{\MEDOC: Meta-clustering Approach}
\MEDOC~  ({\bf Me}ta Clustering based {\bf D}isjoint and {\bf O}verlapping {\bf C}ommunity Detection) starts by executing all base CD algorithms, each with different vertex orderings, to generate a set of community structures. It then creates a multipartite
network. After this, a CD algorithm is used to partition the multipartite network. Finally, a vertex-to-community association function is used to determine the membership of a vertex in a community. Unlike \ENDISCO, \MEDOC~ yields both disjoint and overlapping community structures from the network. 

\subsection{Algorithmic Description}
\MEDOC~ has the following four basic steps (pseudo-code is in Algorithm \ref{meclud}):

\noindent{\bf (i) Constructing multipartite network.} \MEDOC~takes $M$ CD algorithms ${\cal AL}=\{Al_m\}_{m=1}^M$ and runs each $Al_m$ on $K$ different vertex orderings of $G$. For each ordering $k$, $Al_m$ produces a community structure $\mathbb{C}_m^k=\{C_m^{1k},...,C_i^{ak}\}$ of varying size (Step \ref{algo2:perm}). After running on $K$ vertex orderings, each algorithm $Al_m$ produces $K$ different community structures $\Gamma_m=\{\mathbb{C}_m^k\}_{k=1}^K$.  Therefore at the end of Step \ref{algo2:perm}, we obtain $K$ community structures each from $M$ algorithms (essentially, we have $P=M.K$ community structures). We now construct a $P$-partite network (aka meta-network) $GP$ as follows:
the vertices are members of $\bigcup_{m} \mathbb{C}^k_m$, i.e., a community present in a community structure obtained from any of the base algorithms in ${\cal AL}$ and any vertex ordering, is a vertex of $GP$.
We draw an edge from a community $C_m^{ik}$ to a community $C_n^{jk'}$ and associate a weight $W(C_m^{ik},C_n^{jk'})$  (Step \ref{algo2:cons}). Possible definitions of $W$ will be given later in Section~\ref{algo2:Parameter}.
Since each $\mathbb{C}_m^k$ is disjoint, the  vertices in each partition are never connected.   

\noindent{\bf (ii) Re-clustering the multipartite network.} 
Here we run any standard CD algorithm $RAlgo$ on the multipartite network $GP$ and obtain a community structure containing (say) $L$ communities $\mathbb{C}_{GP}=\{C_{GP}^l\}_{l=1}^L$. Note that in this step, we indeed cluster the communities obtained earlier in Step 2; therefore each such community $C_{GP}^l$ obtained here is called a ``meta-community'' (or community of communities) (Step \ref{algo2:run}).   

\noindent{\bf (iii) Constructing an association matrix.} We determine the association between a vertex $v$ and a meta-community $C_{GP}^l$ by using a function $\mathcal{F}$  and construct an association matrix $\mathbb{A}$ of size $|V|\times L$, where each entry ${\mathbb A}(v,l)={\cal F}(v,C_{GP}^l)$ (Step \ref{algo2:asso}). Possible definitions of $\mathcal{F}$ will be given later in Section~\ref{algo2:Parameter}.

\noindent{\bf (iv) Discovering final community structure.} Final vertex-to-community assignment is performed by processing $\mathbb{A}$. The entries in each row of $A$ denote membership probabilities of the corresponding vertex in $L$ communities. For disjoint community assignment, we label each vertex $v$ by the community $l$ in which $v$ possesses the most probable membership in $\mathbb{A}$, i.e.,  $l^* = \argmax\limits_{l} \ \mathbb{A} (v,l)$. Tie-breaking is handled by assigning the vertex to the community to which most of its direct neighbors belong. Note that every meta-community can not be guaranteed to contain at least one vertex, that in turn can not assure $L$ communities in the final community structure. For discovering overlapping community structure, we assign a vertex $v$ to those communities for which the membership probability exceeds a threshold $\delta$. Possible ways to specify threshold will be specified later in Section~\ref{algo2:Parameter}.

\subsection{Parameter Selection}\label{algo2:Parameter}
\noindent $\bullet$ {\bf Matching Function ($W$)}: Given two communities $C_i$ and $C_j$, we measure their matching/similarity via Jaccard Coefficient ({\em JC})=$\frac{|C_i \cap C_j|}{|C_i \cup C_j|}$ and average precision ({\em AP}) =$\frac{1}{2}(\frac{|C_i \cap C_j|}{|C_i|} + \frac{|C_i \cap C_j|}{|C_j|})$.

 \noindent $\bullet$ {\bf Association Function ($\mathcal{F}$)}: Given a meta-community $C$ consisting of (say,) $\gamma$ communities, the association of $v$ with $C$ can be calculated as $\mathcal{F}(v,C)=\frac{\sum_{l=1}^{\gamma} \delta(v,C^l)}{\gamma}$, where $\delta$ returns $1$ if $v$ is a part of $C^l$, $0$ otherwise. For example, if $C=\{\{1,2,3,5\},\{1,2,7\},\{2,7,8\}\}$, then $\mathcal{F}(1,C)=\frac{2}{3}$. Alternatively,
 a weighted association measure may assign a score to $v$ w.r.t. $C$ based on the co-occurrence of the other community members with $v$, i.e., {\scriptsize $\mathcal{F}_w(v,C)=  
\frac{\big|\bigcap \limits_{C^l\in C}  C^l \delta(v,C^l) \big|} {\big|\bigcup\limits_{C^l\in C} C^l \delta(v,C^l)\big|}$}. In our earlier example, $\mathcal{F}_w(1,C)=\frac{|\{1,2\}|}{|\{ 1,2,3,5,7\}|}=\frac{2}{5}$.    

\noindent $\bullet$ {\bf Threshold ($\tau$):} We choose the threshold $\tau$ automatically as follows. We first assign each vertex to its most probable community -- this produces a disjoint community structure. 
Each vertex $v_i$ is represented by a feature vector $F(v_i)$ which is the entire $i$'th row of the association matrix $\mathbb{A}$.
We then measure the average similarity of vertices in $C$ as follows: $AS(C)=\frac{\sum_{(u,v)| u,v \in C \wedge  E_{uv}\in E_C} COS(F(u),F(v))}{|E_C|}$, where $E_C$ is the set of edges completely internal to $C$, $E_{uv}$ is an edge $(u,v)$, and $COS$ is cosine similarity. The probability that two vertices are connected in $C$ is then defined as:
\begin{equation}\small
 P(C)=\frac{e^{{[AS(C)]}^2}}{1+e^{{[AS(C)]}^2}}
\end{equation}
For a vertex $v$, if $P(C\cup \{v\}) \geq P(C)$, we further assign $v$ to $C$, in addition to  its current community.  We compare our threshold selection method with the following method: each vertex is assigned to its top $n\%$ high probable communities (we set $n$ to $5\%$ or $10\%$). Our experiments show that \MEDOC~delivers excellent performance with our threshold selection method (see Figures \ref{parameter_over}(g)-(i)).

Other input parameters $RAlgo$ and $K$ remain same as discussed in Section \ref{algo1:Paramater}.

\vspace{-2mm}

\subsection{Complexity Analysis}
 The most expensive step of \MEDOC~ is to construct the multipartite network in Step 3. If $M$ is the number of base algorithms, $K$ is the number of vertex orderings and $\bar{a}$ is the average size of a base community structure,  the worst case scenario occurs when each vertex in one partition is connected to each vertex in other partitions --- if this happens, the total number of edges is $\mathcal{O}(\bar{a}^2M^2K^2)$. 
 However, in practice the network is extremely sparse and  leads to  $\mathcal{O}(\bar{a}MK)$ edges (because in sparse graphs $\mathcal{O}(|V|)\sim \mathcal{O}(|E|)$).
 Further, constructing the association matrix would take $\mathcal{O}(NL)$ iterations (where $L\ll N$).


\begin{table*}
\caption{Properties of the real-world networks. $N$: number of vertices, $E$: number of edges, $C$: number of
communities, $\rho$: average edge-density per community, $S$: average size of a community, $O_m$: average number of community memberships per vertex.}\label{dataset}
\centering
\scalebox{0.75}{
\begin{tabular}{l||l|l|l|r|r|r|r|r|r|c}
\multicolumn{11}{c}{(a) Networks with disjoint communities}\\\hline
Networks & Vertex type & Edge type & Community type & N & E & C & $\rho$ & S & $O_m$ & Reference\\\hline
University & Faculty      & Friendship          & School        &  81      &  817      &    3    & 0.54    &   27     &  1  & \cite{PhysRevE.77.016107} \\
Football & Team & Games & Group-division  & 115 & 613 & 12 & 0.64  & 9.66 & 1 & \cite{Clauset2004} \\
Railway & Stations & Connections & Province & 301 & 1,224 & 21 & 0.24 & 13.26 & 1 & \cite{Chakraborty:2014} \\  
Coauthorship & Researcher & Collaborations & Research area & 103,677 & 352,183 & 24 &  0.14  & 3762.58 & 1 & \cite{0002SGM14,ChakrabortySTGM13} \\\hline

\multicolumn{11}{c}{}\\
\multicolumn{11}{c}{(b) Networks with overlapping communities}\\\hline
Networks & Vertex type & Edge type & Community type & N & E & C & $\rho$ & S & $O_m$ & Reference\\\hline
Senate & Senate & Similar voting pattern & Congress & 1,884 & 16,662 & 110    & 0.45 & 81.59 & 4.76 & \cite{PhysRevE.91.012821}\\ 
Flickr & User & Friendship & Joined group & 80,513 & 5,899,882 & 171 & 0.046 & 470.83 & 18.96 & \cite{Wang-etal12} \\
Coauthorship & Researcher & Collaborations & Publication venues &  391,526 & 873,775 & 8,493 & 0.231 & 393.18& 10.45 & \cite{Palla}\\
LiveJournal & User & Friendship & User-defined group  & 3,997,962 & 34,681,189 & 310,092 & 0.536  & 40.02 & 3.09 & \cite{Leskovec} \\
Orkut  & User & Friendship & User-defined group & 3,072,441 & 117,185,083 & 6,288,363 & 0.245 & 34.86 & 95.93 & \cite{Leskovec}\\\hline
\end{tabular}}
\vspace{-3mm}
\end{table*}

%
%
%
%

\section{Results of Disjoint Community Detection}

\subsection{Datasets}\label{dis_dataset}
We use the LFR benchmark model \cite{PhysRevE} to generate synthetic networks with ground-truth community structure by varying the number of vertices $n$, mixing parameter $\mu$ (the ratio of inter- and intra-community edges), average degree $\bar k$, maximum degree $k_{max}$, minimum (maximum) community size $c_{min}$ ($c_{max}$), average percentage $O_n$ of overlapping vertices and the average number $O_m$ of communities to which a vertex belongs.\footnote{
Unless otherwise stated, we generate networks with the same parameter configuration used in \cite{Chakraborty:2014,Kanawati2014,1742}: $n=10000$, $\bar k=50$, $k_{max}=150$, $\mu=0.3$, $O_n=0$, $O_m=1$, $c_{max}=100$, $c_{min}=20$.}  
Note that for each parameter configuration,
we generate 50 LFR networks, and the values in all the experiments are reported by averaging the results.
We also use $4$ real-world networks mentioned in Table \ref{dataset}(a) for experiments (see detailed description in Appendix \cite{si}).

\subsection{Baseline Algorithms}
We compare \ENDISCO~and \MEDOC~with the following algorithms:
\ (i) {\em Modularity-based}: FastGreedy (\FG) \cite{newman03fast}, Louvain (\LOU) \cite{blondel2008} and \CNM~ \cite{Clauset2004}; (ii) {\em Vertex similarity-based}: WalkTrap (\WT) \cite{JGAA-124}; (iii) {\em Compression-based}: InfoMap (\INF) \cite{Rosvall29012008}; (iv) {\em Diffusion-based}: Label Propagation (\LP) \cite{raghavan-2007}. These algorithms are also used as base algorithms in ${\cal AL}$ in our ensemble approaches. 
We further compare our methods with Consensus Clustering (\CC) \cite{lanc12consensus}, a recently-proposed ensemble-based framework for disjoint community detection.

\subsection{Evaluation Metrics} \label{dis_eval}
As we know the ground-truth community structure, we measure performance of competing CD algorithms using the standard {\em Normalized Mutual Information} ({\em NMI}) \cite{danon2005ccs} and {\em Adjusted Rand Index} ({\em ARI}) \cite{hubert1985}.

  \vspace{-3mm}
\begin{figure}
\centering
 \scalebox{0.23}{
 \centering
  \includegraphics{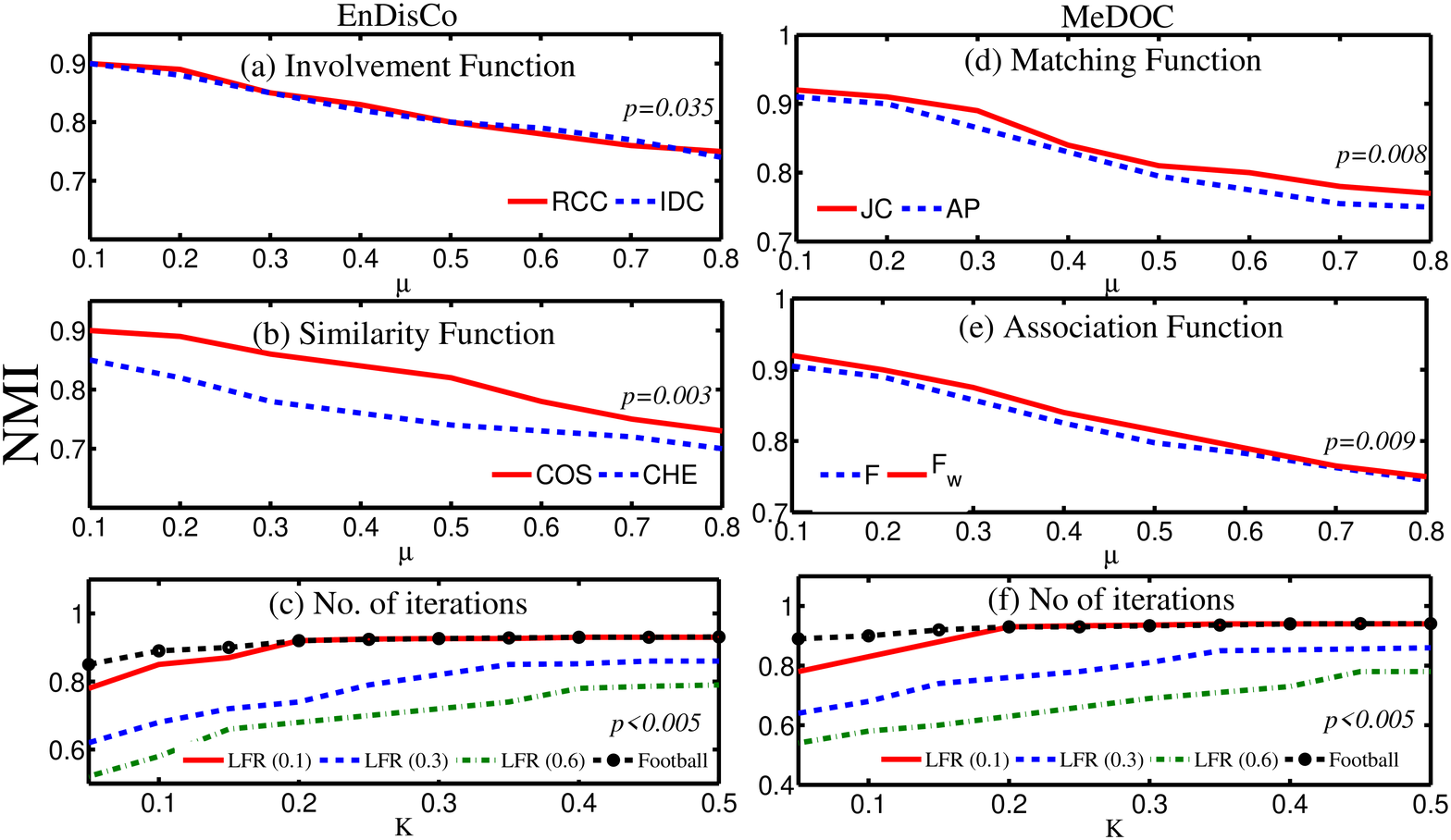}}
    \vspace{-3mm}
  \caption{Dependencies of the performance of \ENDISCO~ (left panel) and \MEDOC~ (right panel) on different parameters. The quality of the ground-truth community is varied by changing $\mu$ from $0.1$ to $0.8$ (keeping the other LFR parameters default) and the performance is measured using NMI. In (c) and (f), we vary $K$ and report the accuracy for three different LFR and Football networks. The value corresponding to one parameter is reported by averaging the values for all possible combinations of the other parameters. The results are statistically significant (for multiple curves in (c) and (f), we report the range of $p$-values).}\label{parameter_dis}
   \vspace{-7mm}
\end{figure}

\begin{table*}[!ht]
\centering
\caption{(A) Relative percentage improvement (averaged over NMI and ARI) of \ENDISCO~(E) and \MEDOC~(M) over the baseline algorithms for disjoint community detection. Each row corresponds to an algorithm $Al$ and the value indicates the performance improvement of the ensemble approach with $Al$ as the re-clustering algorithm over the isolated performance of $Al$ without ensemble.  (B) We further compare the our ensemble methods with \CC~(C) for each base algorithm separately and the results are reported averaging over all the networks. The rows in Table-B are same as Table-A. } \label{performance:non}
\scalebox{0.8}{
\begin{tabular}{|l|cccccccc|c|ccccccccccc|c|c>{\columncolor[gray]{0.8}}cc|}
\cline{1-21} \cline{23-25}
\multirow{2}{*}{Algorithm} & \multicolumn{8}{c|}{{\bf Synthetic Networks}} & &\multicolumn{11}{c|}{{\bf Real-world Networks}} & &\multicolumn{3}{c|}{{{\bf Average over} }}\\
\multirow{2}{*}{ } & \multicolumn{2}{c}{LFR ($\mu=0.1$)} & &\multicolumn{2}{c}{LFR ($\mu=0.3$)} & &\multicolumn{2}{c|}{LFR ($\mu=0.6$)} & &\multicolumn{2}{c}{Football} & &\multicolumn{2}{c}{Railway} & &\multicolumn{2}{c}{University} & &\multicolumn{2}{c|}{Coauthorship} &  & \multicolumn{3}{c|}{{\bf All Networks}}\\ \cline{2-3} \cline{5-6} \cline{8-9} \cline{11-12} \cline{14-15} \cline{17-18} \cline{20-21} \cline{23-25}
                           & E               & M     &         & E               & M &              & E               & M&              & E           & M&           & E           & M&          & E            & M&            & E             & M&& \multicolumn{1}{c}{E} &M & C              \\ \cline{1-9} \cline{11-21}  \cline{23-25}

\FG     & 2.39                     &    2.93   &            &    2.71             &     3.02   &           &        3.81         &  3.92        &         &   0              &       0   &     &   1.22           &   1.43   &         &    2.20         &        2.86   &     &       3.98        &       4.60  & & 2.33 & 2.36 &  1.98             \\                           

\LOU     & 1.97                    &      2.04             &    &  2.22           &     2.40              &     &        3.41    &    3.86          &     & 0                &     0 &          &   1.17           			&        1.43       &   & 2.12          &   2.30 &             &   2.21           &     2.39 & & 1.99 & 2.01 &  1.98           \\
\CNM     &2.07                       &      2.46   &          &   2.14              &    2.83 &   &3.22              &    3.50 &             &   1.23                &   1.46         &     &       1.49        &  1.92            &  & 2.39            &  2.40           &  &    2.92          &    3.41     &   & 2.20 & 2.42 &  2.01 \\

\INF     & 0                          &      0         &    &     1.44            &    1.62           &    &  2.01               &   2.46  &                &  0               &        0 &      &    1.22          &          1.56  &     &     2.01        &        2.20        &               &    2.31             &      2.98 & & 1.28 & 1.31 &   1.28        \\
\WT    &4.43                        &         4.97          &     & 4.86            &    5.08               &     & 6.98            &    7.42    &           &  2.21               &     2.46 &          &    3.21          &    3.49     &      &   4.22          &   4.49             &   &     5.06       &      5.51    & & 4.24 & 4.65 & 4.01      \\

\LP    & \cellcolor{gray!25} 5.06                        &       \cellcolor{gray!25}      5.72 &     &    \cellcolor{gray!25}  5.12           &   \cellcolor{gray!25}  5.39              &    &  \cellcolor{gray!25} 7.50          &   \cellcolor{gray!25}   7.82 &             &      \cellcolor{gray!25} 3.01           &  \cellcolor{gray!25}   3.29 &          &   \cellcolor{gray!25} 3.46      &    \cellcolor{gray!25}  3.79 &          &   \cellcolor{gray!25}  6.21        &  \cellcolor{gray!25} 6.80 &             &   \cellcolor{gray!25}  6.21          &  \cellcolor{gray!25} 6.98     & & 5.21 & 5.46 & 3.76     \\\cline{1-21}  \cline{23-25}
\multicolumn{21}{c}{(A)} &\multicolumn{1}{c}{}& \multicolumn{3}{c}{(B)} 

\end{tabular}}

 \vspace{-8mm}
\end{table*}

\subsection{Experimental Results}
We first run experiments to identify the best parameters for \ENDISCO~and \MEDOC~and then present the comparative analysis among the competing algorithms.

\subsubsection{Dependency on the Parameters}
We consider the LFR networks and vary  $\mu$. Figure \ref{parameter_dis}(a) shows that the accuracy of \ENDISCO~is similar for both the involvement functions, while  Figure \ref{parameter_dis}(b) shows cosine similarity fully dominating Chebyshev distance. Figure \ref{parameter_dis}(d) shows that Jaccard coefficient performs significantly better than average precision when \MEDOC~is considered, while \ Figure \ref{parameter_dis}(e) shows that the weighted association function seems to dominate the other for $\mu < 0.6$ and exhibits similar performance for $\mu\geq 0.6$.  We further vary the number of iterations $K$ to obtain communities with different vertex orderings -- Figures \ref{parameter_dis}(c) and \ref{parameter_dis}(f) show that for the networks with strong community structure (such as LFR ($\mu=0.1$), Football), the accuracy levels off at $K=0.2|V|$; however with increasing $\mu$ leveling off occurs at larger values of $K$. Note that the patterns observed here for LFR network are similar for other networks. Therefore unless otherwise stated, in the rest of the experiment we show the results of our algorithms with the following parameter settings for disjoint community detection: \ENDISCO: $K=0.2|V|$, $RCC$, $COS$; \MEDOC: $K=0.2|V|$, $JC$, $F_w$.

\subsubsection{Impact of Base CD Algorithms on \ENDISCO~and \MEDOC}\label{sec:impact}
In order to assess the  impact of each base algorithm in our ensemble, we measure the accuracy of \ENDISCO~and \MEDOC~when that base algorithm is removed from the ensemble --- Table \ref{impact} shows that for LFR networks \INF~has the biggest impact on accuracy according to both the evaluation measures (NMI and ARI) for both \ENDISCO~and \MEDOC~ (results are same for real networks \cite{si}). 

As the final step in both \ENDISCO~and \MEDOC~is to run a CD algorithm for re-clustering, we also conduct experiments (Table~\ref{impact_den} for LFR networks and Appendix \cite{si} for real networks) to identify the best one. Again, \INF~proves to be the best.

\begin{table}[!h]
 \centering

 \caption{Impact of each base algorithm on the accuracy of \ENDISCO~ and \MEDOC. The results are reported on default LFR network with default parameter settings of the proposed algorithms (we use \INF~ as the final re-clustering algorithm). Each base algorithm is removed in isolation during the construction of ensemble matrix.}\label{impact}
  \scalebox{0.8}{
 \begin{tabular}{|l|l|c|c|c|c|c|c|}
 \hline
 
\multirow{3}{*}{No.} & Base & \multicolumn{4}{c|}{Disjoint} & \multicolumn{2}{c|}{Overlapping} \\\cline{3-8}
              &Algorithm & \multicolumn{2}{c|}{\ENDISCO} & \multicolumn{2}{c|}{\MEDOC} & \multicolumn{2}{c|}{\MEDOC} \\\cline{3-8}
            &  &  NMI & ARI & NMI & ARI & ONMI & $\Omega$ \\\hline
(1) & All & 0.85 & 0.89  & 0.87  & 0.90  & 0.84 & 0.87  \\
(2) & (1) $-$ \FG & 0.83 & 0.88 & 0.84 & 0.88 & 0.83 & 0.85 \\
(3) & (1) $-$ \LOU & 0.82 & 0.86 & 0.85 & 0.86 & 0.81 & 0.84 \\
(4) & (1) $-$ \CNM & 0.82 & 0.85 & 0.83 & 0.87 & 0.82 & 0.85 \\
\rowcolor[HTML]{D3D3D3}
(5) & (1) $-$ \INF & 0.80 & 0.81 & 0.81  & 0.82 & 0.80 & 0.81\\
(6) & (1) $-$ \WT & 0.84 & 0.88 & 0.85  & 0.81 & 0.83  & 0.86\\
(7) & (1) $-$ \LP & 0.84 &  0.87 & 0.86 & 0.87 & 0.83 & 0.85\\\hline
 \end{tabular}}
 \vspace{-5mm}
\end{table}

 
\begin{table}[!h]
 \centering

 \caption{Impact of each algorithm  at the final stage of \ENDISCO~ and \MEDOC~ to re-cluster vertices. The results are reported on default LFR network with other default parameter values of the proposed algorithms.}\label{impact_den}
  \scalebox{0.80}{
 \begin{tabular}{|l|c|c|c|c|c|c|}
 \hline
 
Re-clustering & \multicolumn{4}{c|}{Disjoint} & \multicolumn{2}{c|}{Overlapping} \\\cline{2-7}
Algorithm & \multicolumn{2}{c|}{\ENDISCO} & \multicolumn{2}{c|}{\MEDOC} & \multicolumn{2}{c|}{\MEDOC} \\\cline{2-7}
            &  NMI & ARI & NMI & ARI & ONMI & $\Omega$ \\\hline

\FG  &  0.79 & 0.80 & 0.80 & 0.83 & 0.81 & 0.84\\
\LOU & 0.82 & 0.84 & 0.83 & 0.86 & 0.82 & 0.83\\
\CNM & 0.83 & 0.81 & 0.83 & 0.86 & 0.81 & 0.80\\
\rowcolor[HTML]{D3D3D3}
\INF & 0.85 & 0.89  & 0.87  & 0.90  & 0.84 & 0.87  \\
\WT & 0.75 & 0.78 & 0.77 & 0.82 & 0.76 & 0.79\\
\LP & 0.77 & 0.79 & 0.78 & 0.80 & 0.75 & 0.77\\\hline

 \end{tabular}}
 \vspace{-6mm}
\end{table}

\subsubsection{Comparative Evaluation}\label{comp_dis}
Table \ref{performance:non}(A) reports the performance of our approaches on all networks using different algorithms in the final step of \ENDISCO~and \MEDOC. The numbers denote relative performance improvement of our algorithms (E:\ENDISCO\ M:\MEDOC) w.r.t.
a given algorithm when that algorithm is used in the final step. For instance, the first entry in the last row (5.06) means that for LFR ($\mu=0.1$) network, the accuracy of \ENDISCO~(when \LP\ is used for re-clustering in its final step) averaged over NMI and ARI (0.83) is 5.06\% higher than that of \LP\ (0.79). The actual values are reported in Appendix \cite{si}. The point to take away from this table is that irrespective of which classical CD algorithm we compare against, \ENDISCO~and \MEDOC~always improve the quality of communities found. Moreover, we observe from the results of LFR networks that with the deterioration of the community structure (increase of $\mu$), the improvement increases for all the re-clustering algorithms. Further,
Table \ref{performance:non}(B) shows the average improvement 
of \ENDISCO~and \MEDOC~when compared against Consensus Clustering (\CC). We see that for disjoint networks, both \ENDISCO~and \MEDOC~beat \CC~ with \MEDOC~emerging in top place.



\section{Results of Overlapping Community Detection}

\subsection{Datasets}
We again use the  LFR benchmark to generate synthetic networks with overlapping community structure with the following default parameter settings as mentioned in \cite{oslom,Gregory1}: $n=10000$, $\bar k=50$, $k_{max}=150$, $\mu=0.3$, $O_n=20\%$, $O_m=20$, $c_{max}=100$, $c_{min}=20$.
We generate $50$ LFR networks for each parameter configuration --- the experiments reported averages over these $50$ networks. We further vary $\mu$ ($0.1$-$0.8$ with increment of $0.05$), $O_m$ and $O_n$ (both from $15\%$ to $30\%$ with increment of $1\%$) depending upon the  experimental need.

We also run experiments with six real-world datasets mentioned in Table \ref{dataset}(b) (see details in Appendix \cite{si}).

\subsection{Baseline Algorithms}
We compare \MEDOC~ with the following state-of-the-art overlapping community detection algorithms: 
(i) {\em   Local expansion:} OSLOM \cite{oslom}, EAGLE \cite{Shen}; (ii) {\em Agent-based dynamical algorithms:} COPRA \cite{Gregory1}, SLPA \cite{Xie}; (iii) {\em Detection using mixture model:} MOSES \cite{moses}, BIGCLAM \cite{Leskovec}.

\subsection{Evaluation Metrics}
We use the following evaluation metrics to compare the results with the ground-truth community structure: (a) Overlapping Normalized Mutual Information ($ONMI$) \cite{journals}, (b) Omega ($\Omega$) Index \cite{Leskovec} (details in Appendix \cite{si}).

\begin{table*}[!t]
\begin{center}
\caption{Accuracy of all the competing algorithms in detecting the overlapping community structure from both synthetic and real-world networks. All the disjoint algorithms are used to create  the multipartite network and \MEDOC~ is run with its default parameter setting. } \label{performance:over}
\scalebox{0.7}{
\begin{tabular}{|l|cccccccc|c|cccccccccccccc|c|}
\hline

\multirow{2}{*}{Algorithm} & \multicolumn{8}{c|}{{\bf Synthetic Networks}} & &\multicolumn{14}{c|}{{\bf Real-world Networks}} \\
\multirow{2}{*}{ } & \multicolumn{2}{c}{LFR ($\mu=0.1$)} & &\multicolumn{2}{c}{LFR ($\mu=0.3$)} & &\multicolumn{2}{c|}{LFR ($\mu=0.6$)} & &\multicolumn{2}{c}{Senate} & &\multicolumn{2}{c}{Flickr} & &\multicolumn{2}{c}{Coauthorship} & &\multicolumn{2}{c}{LiveJournal} & &\multicolumn{2}{c|}{Orkut} \\ \cline{2-3} \cline{5-6} \cline{8-9} \cline{11-12} \cline{14-15} \cline{17-18} \cline{20-21} \cline{23-24} 
                           & ONMI               & $\Omega$     &         & ONMI               & $\Omega$ &              & ONMI               & $\Omega$&              & ONMI           & $\Omega$&           & ONMI           & $\Omega$&          & ONMI            & $\Omega$&            & ONMI             & $\Omega$&     & ONMI & $\Omega$  \\ \cline{1-9} \cline{11-24}

\OSLOM                 & 0.80	&0.78&&	0.74&	0.78&&	0.72&	0.73&&	0.71&	0.73	&&0.68&	0.74&&	0.70&	0.71&&	0.73&	0.75&&	0.71	&0.76

      \\                           

\EAGLE     &  0.81	&0.83	&&0.75&	0.76&&	0.70	&0.74&&	0.73	&0.74&&	0.69	&0.76&&	0.71&	0.74&&	0.74&	0.76&&	0.70	&0.77
        \\
\COPRA    & 0.80	&0.81&&	0.76&	0.74&&	0.72&	0.74&&	0.74	&0.77&&	0.73&	0.78&&	0.75&	0.79&&	0.76&	0.82&&	0.74&	0.76
  \\

\SLPA       & 0.84&	0.86&&	0.78&	0.77&&	0.76&	0.77&&	0.74&	0.76&&	0.72&	0.74&&	0.76&	0.77&&	0.78&	0.85&&	0.75&	0.79\\

\MOSES            & 0.85&	0.86	&&0.80&	0.81&&	0.75&	0.78&&	0.75&	0.78&&	0.74&	0.76&&	0.79&	0.78&&	0.81&	0.82&&	0.78&	0.82
   \\

\BIGCLAM         & 0.86	&0.85	&&0.81&	0.83&&	0.77&	0.79&&	0.76&	0.79&&	0.75&	0.76&&	0.80&	0.84&&	0.84&	0.87&&	0.81&	0.84
     \\
\MEDOC         &  \cellcolor{gray!25} 0.88&	\cellcolor{gray!25}0.91	&&\cellcolor{gray!25}0.84&	\cellcolor{gray!25}0.87&&	\cellcolor{gray!25}0.82&	\cellcolor{gray!25}0.84&&	\cellcolor{gray!25}0.81&	\cellcolor{gray!25}0.85&&	\cellcolor{gray!25}0.79&	\cellcolor{gray!25}0.84&&	\cellcolor{gray!25}0.82&	\cellcolor{gray!25}0.86&&	\cellcolor{gray!25}0.86&	\cellcolor{gray!25}0.88&&	\cellcolor{gray!25}0.83&	\cellcolor{gray!25}0.86
  \\

\hline

\hline
\end{tabular}}
 
\end{center}
 \vspace{-8mm}
\end{table*}

\subsection{Experimental Results}

\subsubsection{Parameter Settings}
\begin{figure}
\centering
 \scalebox{0.2}{
  \includegraphics{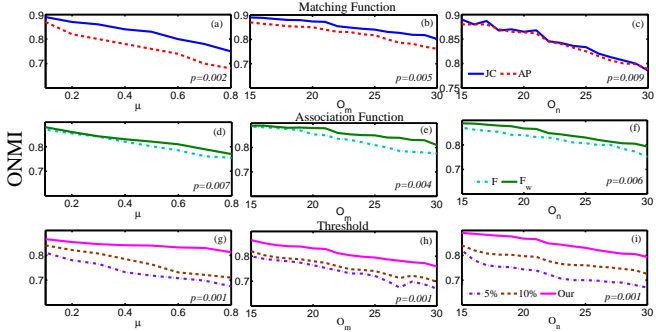}}
  \caption{Dependencies of \MEDOC~ on different algorithmic parameters. The results are reported on default overlapping LFR networks by varying three parameters $\mu$, $O_m$ and $O_n$. For thresholding, we choose top 5\% and 10\% highly probable communities for each vertex and compare it with our threshold selection method.  The value corresponding to one parameter is reported by averaging the values for all possible combinations of the other parameters. The results are statistically significant.}\label{parameter_over}
   \vspace{-8mm}
\end{figure}

We first try to identify the best parameter settings for \MEDOC.  These include: 
matching function $W$, association function $\mathcal{F}$,  number of iterations $K$ and threshold $\tau$. Figure \ref{parameter_over} shows the results (on LFR networks) by varying $\mu$, $O_m$ and $O_n$. 
We observe that Jaccard coefficient as matching function and weighted association measure are better than their alternative. The choice of $K$ is the same as shown in Figure \ref{parameter_dis}(f) -- accuracy levels off at $K=0.2|V|$, and therefore we skip this result in the interest of space. We experiment with two choices of thresholding: top 5\% and 10\% most probable communities per vertex, and compare with the threshold selection mechanism described in Section \ref{algo2:Parameter}. Figures \ref{parameter_over}(g)-\ref{parameter_over}(i) show that irrespective of any network parameter selection, our choice of selecting threshold always outperforms others. As shown in Table \ref{impact_den}, \INF~ seems to be the best choice for the re-clustering algorithm. Therefore, in the rest of the experiments, we  run \MEDOC~with $K=0.2|V|$, $JC$, $F_w$, \INF~ and $\tau$ (selected by our method).

\subsubsection{Impact of Base Algorithms for Overlapping CD}
The impact of the base algorithms on \MEDOC's performance is similar to what we saw in the disjoint CD case. The results in Table \ref{impact} show that accuracy decreases most when we drop \INF~from the base algorithm, followed by \LOU~ and \CNM ~(see more results in Appendix \cite{si}).

\subsubsection{Comparative Evaluation}
We ran \MEDOC~with the default setting on three LFR networks and five real-world networks. The performance of \MEDOC~is compared with the six baseline overlapping community detection algorithms. Table \ref{performance:over} shows the performance of the competing algorithms in terms of ONMI and $\Omega$ index. In all cases, \MEDOC~is a clear winner, winning by significant margins. The absolute average of ONMI ($\Omega$) for \MEDOC~ over all networks is 0.83	(0.86), which is 3.58\% (4.39\%) higher than \BIGCLAM,  5.90\% (7.49\%) higher than \MOSES, 8.31\% (9.19\%) higher than \SLPA, 10.67\% (10.95\%) higher than \COPRA, 13.89\% (12.95\%) higher than \EAGLE, and 14.68\% (15.21\%) higher than \OSLOM. Another interesting observation is that the performance improvement seems to be prominent with the deterioration of community quality. For instance, the improvement of \MEDOC~w.r.t. the best baseline algorithm (\BIGCLAM) is 2.32\% (7.06\%), 3.70\% (4.82\%) and 6.49\% (6.33\%) in terms of ONMI ($\Omega$) with the increasing value of $\mu$ ranging from 0.1, 0.3 and 0.6 respectively. This once again corroborates our earlier observations in Section \ref{comp_dis} for disjoint communities.

\begin{figure}[!h]
\centering
 \scalebox{0.2}{
  \includegraphics{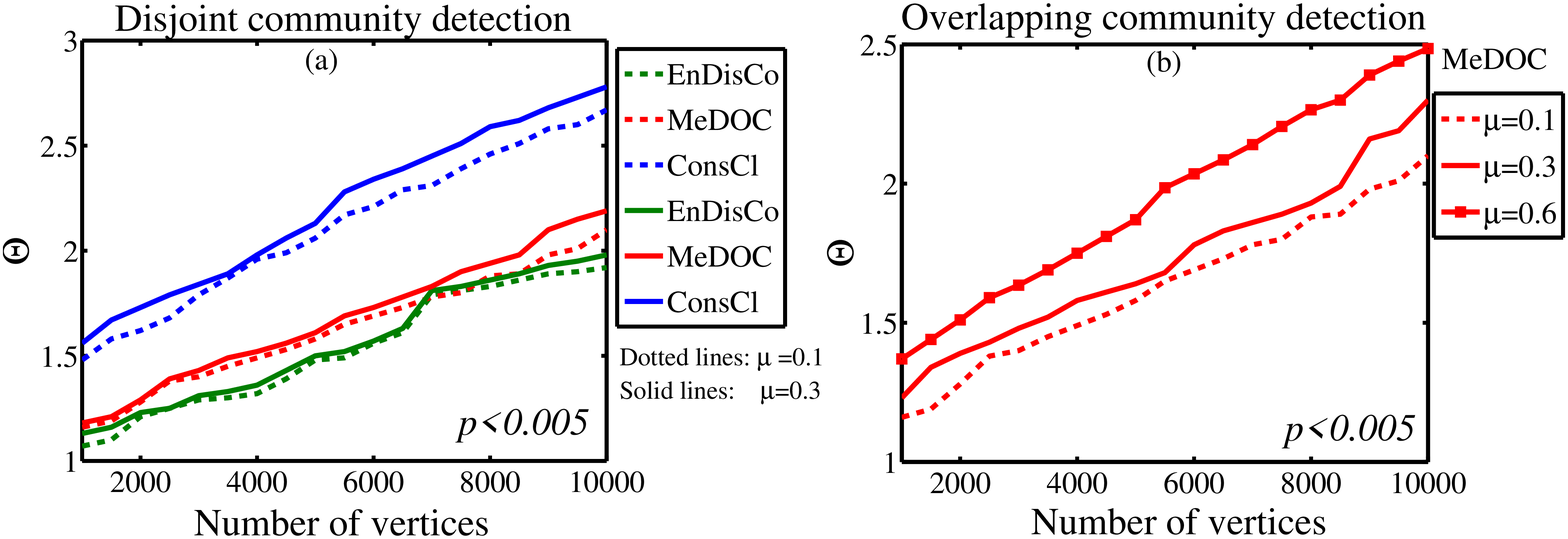}}
  \caption{The value of $\Theta$ w.r.t. the increase of vertices in LFR networks. \ENDISCO~and \MEDOC~ are compared with \CC. The results are statistically significant (since there are multiple curves, we report the range of $p$-values).}\label{runtime}
   \vspace{-3mm}
\end{figure}

\vspace{-3mm}
\section{Runtime Analysis}
Since ensemble approaches require the running all baseline algorithms (which may be parallelized), one cannot expect ensemble methods to be faster than baseline approaches. However, our proposed ensemble frameworks are much faster than existing ensemble approaches such as consensus clustering. To show this, for each ensemble algorithm, we report $\Theta$,
the ratio between the runtime of each ensemble approach and the sum of runtimes of all base algorithms, with increasing number of vertices in LFR. We vary the number of edges of LFR by changing $\mu$ from $0.1$ to $0.3$. Figure \ref{runtime} shows that  our algorithms are much faster than consensus clustering. We further report the results of \MEDOC~for overlapping community detection which is almost same as that of disjoint case since it does not require additional steps apart from computing the threshold.

\vspace{-3mm}
\section{Degeneracy of Solutions}
\begin{figure}[!h]
\centering
 \scalebox{0.205}{
 \centering
  \includegraphics{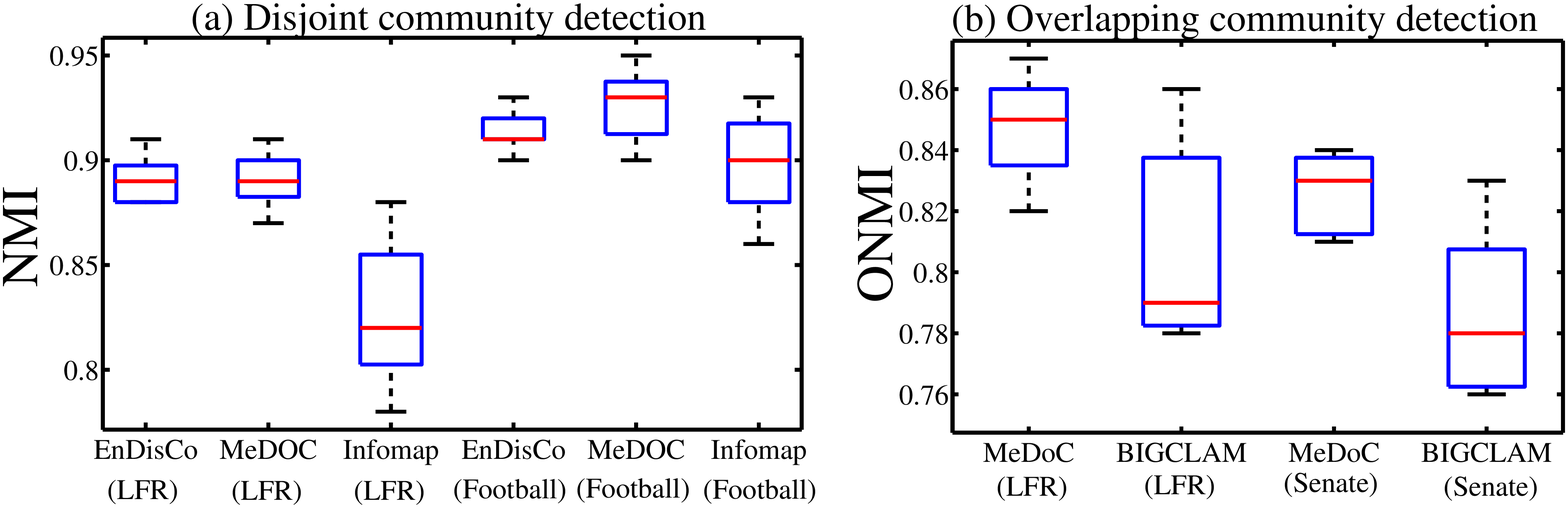}}
  \caption{Box plots indicating the variation of the solutions obtained from \ENDISCO, \MEDOC~ and the best baseline algorithm for (a) disjoint and (b) overlapping CD on one synthetic network and one real-world network.  }\label{boxplot}
   \vspace{-6mm}
\end{figure}

CD algorithms suffer from the problem of ``degeneracy of solutions'' \cite{good2010} which states that an optimization algorithm can produce exponentially many solutions with (nearly-)similar optimal value of the objective function (such as modularity); however the solutions may be structurally distinct from each other. Figure \ref{example} showed how \INF~ produces many outputs for different vertex orderings of Football network.  
We test this by considering the default LFR network and one real-world network (Appendix \cite{si} shows results on more real world networks)
and run the algorithms on $100$ different vertex orderings of each network. We then measure the pair-wise similarity of the solutions obtained from each algorithm. The box plots in Figure \ref{boxplot} show the variation of the solutions for \ENDISCO, \MEDOC~and the 
best baseline algorithm in both disjoint and overlapping community detections. We observe that the median \ similarity is high with \ENDISCO~and \MEDOC~ and the variation is comparatively small. These results suggest that our algorithms provide more robust results than past work and alleviate the problem of degeneracy of solutions.

  \vspace{-3mm}


\section{Conclusion}
In this paper, we proposed two general frameworks for ensemble community detection. \ENDISCO~identifies disjoint community structures, while \MEDOC~detects both disjoint and overlapping community structures. We tested both algorithms on both synthetic data using the LFR benchmark and with several real-world datasets that have associated ground-truth community structure. We show that both \ENDISCO~and \MEDOC~are more accurate than existing CD algorithms, though of course, \ENDISCO~and \MEDOC~leverage them.
We further show that for disjoint CD problems, \ENDISCO~and \MEDOC~both beat a well known existing disjoint ensemble method called consensus clustering \cite{lanc12consensus} -- both in terms of accuracy (measured via both Normalized Mutual Information and Adjusted Rand Index) and run-time. To our knowledge, \MEDOC~is the first ensemble algorithm for overlapping community detection that we have seen in the literature. 
In future, we would like to develop theoretical explanation to justify the superiority of ensemble approaches compared to the discrete models. Other future direction could be to make the ensemble frameworks  parallelized. We will apply the proposed methods to identify communities in specific datasets, such as malware traces, protein interaction networks etc. 

\nop{
In this paper, we proposed two frameworks to aggregate multiple community structures obtained by running different disjoint CD algorithms. Both the frameworks turned out to be superior than past non-ensemble based approaches as well as a recently proposed ensemble approach. We showed how one can leverage disjoint community information  to discover the overlap in the community structure. We presented the suitable functions needed in the process of ensemble and showed an automated way of selecting the threshold. In that sense, our algorithms do not require manual intervention for parameter tunning. 
As it turned out, the choice of base algorithm has no major impact on the clustering quality, but it does in selecting an algorithm for re-clustering.

 In future, we would like to develop theoretical explanation to justify the superiority of ensemble approaches compared to the discrete models.
In traditional ensemble-based classification problem, it has already been shown that an aggregation of several weak base algorithms performs well even though each weak algorithm is just merely better than the random guessing,  i.e., the probability of correct classification is $\frac{1}{|C|} + \epsilon$, where $\epsilon$ is a small non-zero  value and $|C|$ is the number possible classes   \cite{Freund:1995:BWL:220262.220446,Schapire:1990:SWL:83637.83645}. In case of community detection, one can think of weak base algorithms that find the correct community for a vertex in the probability of $\frac{1}{|C|} + \epsilon$ (boosting condition).  After a series of boosting procedures, e.g., boosting by a majority voting, the small $\epsilon$ value would help the probability that the community structure is correctly identified converging to a much larger value than $\frac{1}{|C|} + \epsilon$.  Of course, there exists a certain probability that the majority vote may not work correctly. If the boosting condition is guaranteed for every vertex and the number of weak algorithms is enough, taking a majority vote may produce stable results. In many practical cases, however, we found that the community detection for a subset of vertices is significantly inaccurate by a majority of weak algorithms, and thus the boosting condition might not be guaranteed for all vertices. We leave this line of research as future agenda.
}

\vspace{-3mm}

\section*{Acknowledgment}
Parts of this work were funded by ARO Grants W911NF-16-1-0342, W911NF1110344, W911NF1410358, by ONR Grant N00014-13-1-0703, and Maryland Procurement Ofﬁce under Contract No. H98230-14-C-0137.

\vspace{-3mm}


\end{document}